\pdfoutput=1

\documentclass[11pt]{article}

\usepackage[final]{nlp4MusA}

\usepackage{times}
\usepackage{latexsym}

\usepackage[T1]{fontenc}

\usepackage[utf8]{inputenc}

\usepackage{microtype}

\usepackage{inconsolata}

\usepackage{graphicx}
\usepackage{amsmath}
\usepackage{booktabs}
\title{Harnessing High-Level Song Descriptors towards Natural Language-Based Music Recommendation}

\author{Elena V. Epure, Gabriel Meseguer Brocal, Darius Afchar, Romain Hennequin \\
    Deezer Research, Paris, France \\ \texttt{research@deezer.com}}

\begin{document}
\maketitle
\begin{abstract}
Recommender systems relying on Language Models (LMs) have gained popularity in assisting users to navigate large catalogs.
LMs often exploit item high-level descriptors, i.e. categories or consumption contexts, from training data or user preferences. 
This has been proven effective in domains like movies or products.
However, in the music domain, understanding how effectively LMs utilize song descriptors for natural language-based music recommendation is relatively limited.
In this paper, we assess LMs effectiveness in recommending songs based on user natural language descriptions and items with descriptors like genres, moods, and listening contexts.
We formulate the recommendation task as a dense retrieval problem and assess LMs as they become increasingly familiar with data pertinent to the task and domain.
Our findings reveal improved performance as LMs are fine-tuned for general language similarity, information retrieval, and mapping longer descriptions to shorter, high-level descriptors in music.

\end{abstract}

\section{Introduction}

Music recommender systems are often used to assist users in navigating the vast catalogs offered by streaming platforms.
Recently, Language Models (LMs) have demonstrated efficacy in recommending items such as songs, movies, or books \cite{sanner2023large,penha2020what}. 
They achieve this by 1) matching concise user profiles derived from item preferences expressed in natural language against the available set of items;
and 2) drawing upon their knowledge about specific items gained during pre-training or after fine-tuning for the specific task or domain.

A common thread among these strategies is the reliance of LMs on item high-level descriptors, such as genres (e.g. "romantic comedy" for a movie) or consumption contexts (e.g. "running" for music), within the pre-training corpus or during user preference elicitation phase. 
Empirical evidence suggests that these descriptors play a crucial role in generating relevant recommendations  with LMs, in cases like near cold-start or exploration in movie or product domains \cite{sanner2023large,lopez2021augmenting,malkiel-etal-2020-recobert}.
Leveraging these descriptors has been proven particularly valuable and could surpass the importance of item examples alone such as in collaborative filtering \cite{penha2020what,sanner2023large}.   

In music however, our understanding of how efficiently LMs utilize item textual features and user preference description for music recommendation is comparatively limited.
Previous research has emphasized the significance of natural language features such as tags in improving music retrieval \cite{doh2023toward,laionclap2023} and captioning algorithms \cite{gabbolini-etal-2022-data,doh2023lp}. 
These accessible descriptors help bridge the semantic gap between audio and more complex song descriptions provided by humans \cite{celma2006bridging}. 
Yet, such a study is lacking in the field of music recommendation despite advancements in related areas like conversational recommendation systems \cite{chaganty2023beyond,jannach2021survey,gupta-etal-2023-conversational,jin2019musicbot}.

In this work, we study the efficacy of LMs for song recommendation when users express their preferences in natural language and items are associated with high-level descriptors such as music genres, styles, moods, and listening contexts.
As no large dataset linking natural language user preferences to music descriptors is available, we re-purpose an existing dataset originally created for music captioning \cite{doh2023lp}. 
Subsequently, we formulate the recommendation task as a dense retrieval problem \cite{wang2022gpl} and propose evaluating the LMs as they become increasingly familiar with data pertinent to the target task (recommendation as retrieval) and domain (music). 
This contrasts with previous approaches to integrating LMs, that typically involve encoding low-level item data (such as audio or embedded metadata) and user queries \cite{doh2023toward,laionclap2023}, and are likely less effective because of the inherent semantic gap in music.

Our results show that
a) pre-trained LMs with no further fine-tuning perform quite poorly on this task in music;
b) the performance gets better when the LMs are progressively fine-tuned for text similarity, for multi-domain query retrieval and for mapping longer descriptions on shorter, high-level descriptors in music.
Our approach renders the dataset and models suitable also for retrieving explanations based on song high-level descriptors or encoding any type of item textual information in multi-modal music systems. 
We release the code and fine-tuned models at \url{https://github.com/deezer/nlp4musa_melscribe}.

\section{Related Work}
\label{sec:relatedwork}
Pre-trained LMs are widespread and their usage spans various tasks, including generating synthetic data, producing system utterances, and providing recommendations.
\citet{penha2020what} fine-tune BERT on recommendation conversations on movies, books, or music extracted from Reddit and assess the resulting model's capability to retrieve the most relevant utterance containing a suitable recommendation from existing conversations.
Similarly, \citet{chaganty2023beyond} model the recommendation task as retrieval and fine-tune multiple checkpoints of a T5 encoder to create a common embedding space of conversations and song metadata such (i.e. title, artist and album).
\citet{mysore2023large} build a new dataset by using InstructGPT to generate narrative-driven recommendations in the Point-of-Interest (PoI) domain.
Then, they fine-tune dense retrieval models for this domain.

Other studies depend on LMs for generating system inputs in natural language.
\citet{hayati2020inspired} build a dialogue model comprising two distinct language modules tailored to the recommendation seeker and the recommender system.
\citet{kostric2021soliciting} fine-tune a T5 for learning to generate relevant indirect questions about the context of item consumption in order to help the user to elicit preferences when these are not clearly defined.

Similar to these works, we adopt a recommendation as retrieval approach and assess LMs for our song recommendation task in scenarios where music preferences are given in natural language.
However, our recommendation module relies solely on high-level descriptors rather than low-level ones extracted from audio or embedded metadata (e.g. song titles or artists).
Our hypothesis is that as the model needs to bridge a narrower semantic gap between these two music information sources, it would yield improved outcomes.

\section{Method}
\label{sec:method}
The recommendation task we consider is the learning to rank setup: given $r$, a recommendation request in natural language, and $\mathcal{S}$, a corpus of $N$ songs where each song $s$ has associated $T_s, |T_s| \geq 1$, a set of high-level descriptors (words or short phrases describing music), the recommendation task is addressed with a ranking function $\rho$ over the collection $\mathcal{S}$.
If we consider an encoder LM with an embedding function $f$, which takes as input text and outputs a vector, then the ranking could be computed based on the dot-product similarity between the embeddings of the request $r$ and the concatenated, and alphabetically sorted, high-level descriptors of each song $T_s$:
\begin{math}
\rho(r,\mathcal{S})=(f(r)^\top f(\text{concat}(T_{s_1})),..., f(r)^\top f(\text{concat}(T_{s_N})))    
\end{math}.
The problem revolves around having an efficient text encoder $f$ in music for embedding high-level descriptors and recommendation requests. 

For this, we train a music text bi-encoder using the Generative Pseudo-labeling method (GPL) \cite{wang2022gpl}.
\textit{Bi-encoders} \cite{reimers2019sentencebert} rely on two siamese encoders, comprising a pre-trained LM such as \texttt{BERT} \cite{devlin2019bert} followed by a pooling layer (e.g. \texttt{mean}).
To adapt it to music, we fine-tune it using contrastive learning, which entails that similar input texts (song high-level descriptors on one side and natural language recommendation requests on the other side) are embedded close together in the vector space, while dissimilar ones  are far apart.

Two steps are essential in this method: \textit{the automatic negative mining} and \textit{the pseudo-labeling of training examples}. 
We mine hard negatives $(r, T_s^{-})$ using pre-trained bi-encoders as follows: for each query $r$ we retrieve the top $K$ most similar high-level descriptors, $T_k, k \neq s$.
However, sometimes negatives could be closer to an actual positive example or even a false negative. 
For example, there might be instances when a particular user request for recommendation matches other songs as well, but this was not explicitly labeled in the training data.
As in GPL \cite{wang2022gpl}, we resort to soft-labeling the training data, instead of considering a negative as a true negative.
For each tuple $(r, T_s, T_s^{-})$, we compute a margin score $\delta_s = g(r, \text{concat}(T_s)) - g(r, \text{concat}(T_s^{-}))$ with a cross-encoder $g$ as a teacher.
Compared to bi-encoders that separately encode input texts to a shared vector space, \textit{cross-encoders} take as input the concatenated texts and produce as output a similarity score. 
Cross-encoders are known to be more effective at the text similarity task than bi-encoders but do not scale well \cite{thakur2021augmented}.

\paragraph{Implementation Details}
We choose models pre-trained on \texttt{ms-marco}, a dataset with search queries and documents from various domains \cite{chen2015microsoft}, because of its similarity to our domain and task, that proved useful experimentally too.
We fine-tune \texttt{msmarco-bert-base-dot-v5}\footnote{The name of the used pre-trained models reflects the training dataset (\texttt{msmarco}), the base text encoder (\texttt{bert-base}) and the text similarity function used in training (\texttt{dot})} on our music data  for 1 epoch and $140K$ steps with a batch size of $4$.
This model is chosen as the backbone of our bi-encoder as it yields the best performance in our experiments.
For each $r$, we use \texttt{msmarco-distilbert-base-v3} and \texttt{msmarco-MiniLM-L-6-v3} to mine $30$ negative examples.
To soft-label the training data, we fine-tune a domain-specific cross-encoder on MusicCaps \cite{agostinelli2023musiclm}, a dataset with human-created song captions and descriptors.

\paragraph{Other Baselines}
\texttt{BERT} \cite{devlin2019bert} and \texttt{MPNET} \cite{song2020bert} are pre-trained on general language corpora using various objectives such as masked language modeling or permuted language.  
We apply mean as pooling function to all token embeddings to derive a fixed-size embedding for the given input.
Then, we consider the best-performing bi-encoders reported on \texttt{sentence-transformers} \cite{reimers2019sentencebert}.
Other baselines we include are text encoders obtained from multi-modal (audio-text) representation learning \cite{laionclap2023,doh2023toward}. 
Their text encoding branch is initialized with BERT or RoBERTa weights. 
One difference between \texttt{TTMR} \cite{doh2023toward} and \texttt{CLAP} \cite{laionclap2023} lies in the training dataset and the audio encoding branch.
A \texttt{tf-idf} sparse representation is also considered in the experiments.
Although such a text encoding does not generalise to new vocabulary, we expect it to work well when recommendation requests and music descriptors have high exact term overlap.

\section{Datasets and Evaluation Details}
\label{sec:dataset}
Since there is no large dataset linking natural language user preferences ($r$) with high-level song descriptors ($T_s$) in music recommendation, we re-purpose a dataset originally created for music captioning, LP-MusicCaps \cite{doh2023lp}, to fine-tune our music text bi-encoder.
LP-MusicCaps was created from three pre-existing datasets of audio tracks annotated with tags (the ECALS subset of the Million Song Dataset (\texttt{MSD}) \cite{doh2023toward}, MusicCaps (\texttt{MC}) \cite{agostinelli2023musiclm}, and Magnatagtune (\texttt{MTT}) \cite{law2009evaluation}) by using an instruction-based LLM to generate captions from the given high-level descriptors.

Our goal is to extract training pairs $(r, T_s)$ from LP-MusicCaps by ensuring the compatibility with the desired use case of the text encoder, for conversational music recommendation.
Compared to narrative-driven recommendations, user requests are unlikely \textit{long} in synchronous conversations   \cite{chaganty2023beyond,mysore2023large}.
As the length of each caption in LP-MusicCaps ranges from a sentence to a paragraph, we first split paragraphs in sentences.
Then, as in GPL where each query is seen with multiple documents during training, we ensure multiple sets of high-level descriptors for each sentence.
We sample up to $3$ variations of high-level descriptors from the original $T_s$: first from overlapping high-level descriptors (i.e. tags or phrases from $T_s$ \textit{found} in the sentence) and then from the non-overlapping ones (i.e. tags or phrases from $T_s$ \textit{not} \textit{found} in the sentence).
Like this we simulate cases where there is a varying number of high-level descriptors per song, and some may be irrelevant to the description.

\begin{table*}
\begin{small}
\centering
  \begin{tabular}{l|r|rr|rrr|r}
    \toprule
    & \texttt{Tf-Idf} & \texttt{CLAP}$_{text}$ & \texttt{TTMR}$_{text}$ & \texttt{BERT} & \texttt{all-MiniLM}
    & \texttt{msmarco-BERT} & \texttt{Ours} \\
    \midrule
    MTT & 57.7 $\pm$ 0.8 & 13.5 $\pm$ 0.3 & 7.8 $\pm$ 0.6 & 4.8 $\pm$ 0.4 & 33.3 $\pm$ 0.6 & 32.1 $\pm$ 0.2 & \textbf{62.8 $\pm$ 0.5} \\
    MSD & 30.6 $\pm$ 2.3 & 3.4 $\pm$ 0.1 & 5.1 $\pm$ 0.1 & 4.5 $\pm$ 0.0 & 19.5 $\pm$ 0.2 & 20.7 $\pm$ 0.1 & \textbf{47.9 $\pm$ 0.3}  \\
    MC & \textbf{89.4 $\pm$ 0.4} & 36.5 $\pm$ 1.1 & 19.9 $\pm$ 0.2 & 24.3 $\pm$ 0.9 & 59.9 $\pm$ 0.9 & 66.1 $\pm$ 0.2 & 84.8 $\pm$ 0.2 \\
     MC$_{reco}$ & \textbf{77.7 $\pm$ 0.5} & 27.6 $\pm$ 0.4 & 17.9 $\pm$ 1.2 & 16.3 $\pm$ 0.1 & 48.3 $\pm$ 0.4 & 50.7 $\pm$ 1.0 & 70.1 $\pm$ 0.4 \\
    \bottomrule
  \end{tabular}
  \caption{Recall@10 (mean $\pm$ std) of the all baselines on the LP-MusicCaps test splits.}
\label{tab:evaluation_encoder}
  \end{small}
\end{table*}

In conversational requests for music recommendations, we could find high-level descriptors $T_s$ similar to those in LP-MusicCaps \cite{chaganty2023beyond}.
However, the user utterance $r$ might be formatted like a \textit{request} instead of a statement as in a song caption \cite{jannach2021survey,chaganty2023beyond}.
While we could rephrase each sentence in \texttt{train} as a request, this step is costly and might prove unnecessary. 
The semantic similarity between the two forms of the song description (request versus statement) and the same set of high-level descriptors, $T_s$, might be comparable as it likely relies on topical cues.
In order to check our hypothesis, we rephrase the single-sentence captions from MusicCaps \texttt{test} split from statements into requests for music recommendation with Llama3\footnote{\url{https://ai.meta.com/blog/meta-llama-3/}} (more details in Appendix \ref{app:rephrased}).

We use only the MSD and MTT \texttt{train} splits of LP-MusicCaps for training and keep MC completely unseen.
Each song $s$ is associated with a concatenated set of high-level descriptors $T_s$.
The mapping from the description $r$ to the most likely $T_k, k\in S$ is the proxy for recommending the song $k$.
Then, we compute Recall@10 at the level of descriptors ($T_s$), and not at the song level ($s$) as multiple songs could have the same set of high-level descriptors $T_s$ and $T_s$ is the only song information that we consider in this work.
For each dataset, we produce $3$ test sets by sampling a different set of high-level descriptors per song $s$ and description $r$.
We then report mean and standard deviation (std).

\section{Results}
\label{sec:results}
Table \ref{tab:evaluation_encoder} presents the results obtained from the evaluation of the proposed music text encoder (\texttt{Ours}) and the baselines on song recommendation.
We could notice that \texttt{tf-idf} is a strong encoding function on these datasets where there is a large exact term overlap between the song description $r$ and the high-level descriptors $T_s$.
Though, on the MTT and MSD datasets where this happens less frequently (see Table \ref{tab:stats} for the ratio of descriptor words found in the song description), \texttt{tf-idf}, although competitive, falls short.
Pre-trained LMs such as BERT achieve poor results, most likely because high-level descriptors being short have insufficient context to derive meaningful embeddings\footnote{Similar results were obtained for \texttt{MPNet}.}. 
Although, when used as part of bi-encoders and fine-tuned on a relevant text similarity task (information retrieval), we could see the performance increasing: \texttt{msmarco-BERT} is the best dense retrieval model from the \texttt{sentence-transformers} collection;
we also report results for the second-best, \texttt{all-MiniLM}. 
Similarly, fine-tuning \texttt{BERT} (or variants) on text-audio similarity seem to lead to better text embeddings (see scores for \texttt{CLAP}$_{text}$ and \texttt{TTMR}$_{text}$).
Yet, their performance is less good when exploiting high-level descriptors instead of audio. 

\begin{table}
\begin{small}
\centering
  \begin{tabular}{l|rrr}
    \toprule
  \texttt{test}  & \#Requests & \#Descriptors & Shared Words \\
    \midrule
    MTT & 4462 & 188 & 0.15 \\
    MSD & 34631 & 1054 & 0.23 \\
    MC & 2357 & 6930 & 0.41\\
    MC$_{reco}$ & 2357 & 6930 & 0.34 \\
    \bottomrule
  \end{tabular}
  \caption{Number of requests, unique descriptors, and mean ratio of shared words between each pair ($r$, $T_s$) in \texttt{test}. It could be noticed that in the MC dataset, 40\% of descriptor words are found in the description / request.}
\label{tab:stats}
  \end{small}
\end{table}

The fine-tuned bi-encoder achieves significantly higher scores than all the other dense retrievers.
Compared to $tf-idf$, it does not depend on a pre-established vocabulary.
Thus, by design, it should generalise to new high-level descriptors and is more robust to synonyms and language variations.

Baselines' scores on the rephrased MC test set are lower compared to those on the original MC dataset.
Manual checks revealed that rephrasing descriptions as requests sometimes omitted initial descriptors, thus making it difficult to distinguish between the effects of rephrasing and information loss (a couple of examples are shown in Table \ref{tab:examplesinfolosst} in Appendix \ref{app:rephrased}).
Finally, a more detailed qualitative analysis is presented in Appendix \ref{app:qualitative}. 

\section{Conclusion}
\label{sec:conclusion}
Conversational recommender systems or interfaces based on natural language have emerged as a practical alternative to dynamically elicit preferences from the users in cold-start or exploration cases.
LMs have emerged as central to these systems.
In this work, we analysed the efficacy of LMs towards song recommendation when users express their preferences in natural language and items have high-level descriptors.
We showed that a bi-encoder fine-tuned in multiple phases first for the task and then for the domain is quite competitive.
Future works aims at improving the model on out-of-distribution data, integrating more specialized music knowledge and personalisation during its fine-tuning,
and joining the proposed encoder with other modalities for song recommendation.

\section{Ethical Considerations}

The fine-tuned models, which will be released, target only English-language content and have been exposed primarily to music descriptions and descriptors that mostly refer to Western-centered music, with a limited number of  music descriptions and descriptor set pairs.
Additionally, we are aware that music descriptors such as mood or genre can be specific to individuals, groups, or cultures. 
However, the embeddings we obtain with the fine-tuned models are deterministic and do not take into account any form of localization or personalization, which is a limitation with ethical implications.

\bibliography{nlp4MusA}

\appendix

\section{Rephrased Song Descriptions}
\label{app:rephrased}

In order to rephrase the single-sentence song descriptions from the  MC dataset, we rely on \texttt{meta-llama/Meta-Llama-3-8B-Instruct}, which we prompt as follows:
\newline

\textit{Rephrase this as a music recommendation
}

\textit{request from a user: <<original song description>>}

\textit{Do not use greetings, thanks or emojis.}

\textit{Keep it short, preferably single-sentence.}

\textit{Output:}
\newline

\noindent Each model generation request has a randomly initialised temperature from the list of values \{0.80, 0.85, 0.90, 0.95\}; randomly initialised top\_p value from the same list of values as the temperature; and a randomly initialised seed with a value between 0 and 99999999.
Examples are given in Table \ref{tab:rephrased}.

\begin{table*}
\begin{small}
\centering
  \begin{tabular}{p{7cm}|p{7cm}}
    \toprule
    \textbf{LP-MusicCaps Song Description} & \textbf{ Equivalent Recommendation Request} \\     \midrule
    This amateur recording features a steeldrum melody in a higher register creating a joyful and tropical atmosphere. & I'd love to hear some more upbeat tropical music with a similar steeldrum melody in a higher register.  \\\midrule 
    This is a beautiful folk song, embodying the traditional feel of a middle eastern song, featuring a powerful male voice accompanied by the rhythmic beats of the darbuka and the haunting melody of the oud instrument at a moderate tempo, surrounded by other enchanting middle eastern instruments. & I'm looking for music that sounds like a traditional Middle Eastern folk song with a powerful male vocalist accompanied by darbuka and oud at a moderate tempo.\\ \midrule 

    This fingerstyle-guitar track features delicate acoustic guitar melodies played at a medium tempo, creating a classical atmosphere that is both emotional and soothing. & Recommend a fingerstyle guitar track with a classical atmosphere, featuring delicate acoustic guitar melodies played at a medium tempo, perfect for evoking emotions and providing a soothing listening experience. \\ \midrule
    This k-pop love song features a male vocalist singing in Korean with a youthfully sentimental tone, set to a melodic and dulcet medium tempo track infused with world music influences, including atmospheric synths and chimes, a romantic piano, steady drumming and straightforward bass lines, all backed by a boy band chorus for a pleasantly emotional and ambient experience. & Can you recommend a K-pop love song with a youthful sentimental tone, featuring a male vocalist singing in Korean over a melodic medium tempo track with atmospheric synths, chimes, piano, and boy band chorus? \\
        \bottomrule
  \end{tabular}
  \caption{LP-MusicCaps song descriptions and the equivalent request for music recommendation rephrased with the \texttt{meta-llama/Meta-Llama-3-8B-Instruct} model.}
\label{tab:rephrased}
  \end{small}
\end{table*}

Examples of information loss regarding the descriptors when rephrasing song descriptions as requests are presented in Table \ref{tab:examplesinfolosst}.

\begin{table*}
\begin{small}
\centering
  \begin{tabular}{p{6cm}|p{5cm}|p{3cm}}
    \toprule
    \textbf{Original Song Description} & \textbf{Rephrased Song Request} & \textbf{Music Descriptors} \\ \midrule
    This amateur recording features a steeldrum melody in a higher register creating a joyful and tropical atmosphere. & I'd love to hear some more upbeat tropical music with a similar steeldrum melody in a higher register. & steeldrum, higher register, \underline{amateur recording} \\ \midrule

    This cinematic masterpiece features a blend of haunting sound effects and triumphant horn honking that transports the listener on a thrilling journey through soundscapes. & I'm looking for music that blends haunting sound effects with triumphant horn honking to create an immersive and thrilling soundscape. & \underline{cinematic}, soundeffects, horn honking\\
        \bottomrule
  \end{tabular}
  \caption{Underlined descriptors are no longer mentioned in the rephrased form of song descriptions as requests.}
\label{tab:examplesinfolosst}
  \end{small}
\end{table*}

\section{Qualitative Analysis}
\label{app:qualitative}

In Table \ref{tab:examples}, we present multiple examples of song descriptions from the LP-MusicCaps data, together with their ground-truth associated song descriptors and top five predictions by our model.
In the first row, none of the top 5 sets of descriptors matches exactly the ground-truth; however, most retrieved descriptor sets are still relevant.
In the next examples, the ground-truth descriptor set is found among those retrieved by the model.
However, we could also notice that sometimes irrelevant descriptor sets are returned such as "soft piano" in the forth example, or "exotic" and "amateur recording, r\&b" in the last example.  

\begin{table*}
\begin{small}
\centering
  \begin{tabular}{p{4.5cm}|p{3.5cm}|p{6cm}}
    \toprule
    \textbf{Song Description / Request} & \textbf{Music Descriptors} \newline (Ground-truth) & \textbf{ Top 5 Predictions by \texttt{Ours}} \\
    \midrule
    This heartfelt ballad showcases a soulful and sad low-quality sustained strings melody intertwined with a mellow piano melody, and a soft female vocal, resulting in an emotionally charged and sonically rich experience for listeners. & low quality, sad, sustained strings melody, ballad, mellow piano melody & [1] soft piano \newline [2] low quality, emotional female vocal, mellow piano melody, live performance, r\&b \newline [3] emotional, low quality, reverberant female vocal, sad acoustic rhythm guitar chord progression, soft rock \newline [4] soft \newline [5] sad \\ \midrule
    This pop song features a captivating teen female vocal delivering melodic singing over an acoustic guitar and simple drum track that evoke a melancholic, emotional vibe. & acoustic guitar, emotional, teen female vocal, melodic singing, simple drum track, pop music & [1] \textit{acoustic guitar, emotional, teen female vocal, melodic singing, simple drum track, pop music} \newline
    [2] acoustic guitar \newline
    [3] acoustic drums \newline
    [4] bass drum \newline
    [5] pop, acoustic rhythm guitar, quiet playback, resonant, heartfelt, noisy, emotional, passionate female vocal \\ \midrule
    This song is a perfect blend of country and pop with a touching singer-songwriter flair, featuring an emotional, soulful male voice that's accompanied by the soft strums of an acoustic guitar. & emotional, male singer, country / pop / singersongwriter & [1] soft \newline [2] acoustic guitar \newline [3] \textit{emotional, male singer, country / pop / singersongwriter} \newline [4] male voice \newline [5] soft piano \\ \midrule
    Experience an otherworldly journey through an amateur recording filled with out-of-this-world digital sounds, a hair-raising riser, and a hauntingly atmospheric vibe. & atmospheric & [1] amateur recording \newline  [2] \textit{atmospheric} \newline  [3] digital sound effects, amateur recording \newline [4] exotic \newline [5] amateur recording, r\&b \\
    \bottomrule
  \end{tabular}
  \caption{Song descriptions, their ground-truth descriptor sets, and top 5 predicted descriptor sets by \texttt{Ours}.}
\label{tab:examples}
  \end{small}
\end{table*}

\end{document}